\documentclass[12pt]{article}
\usepackage{bm,latexsym,amsmath,amsfonts,amssymb,fancyhdr,color,graphicx,wasysym}
\usepackage[a4paper,bottom=2.5cm,top=3.cm,head=5mm, left=2.5cm,width=18cm,dvipdfm]{geometry}
\usepackage[usenames,dvipsnames,svgnames,table]{xcolor}
\usepackage[ps2pdf,
            colorlinks=true,
            linkcolor=blue,
            urlcolor=blue,
            citecolor=green,          
            bookmarks=true,
            bookmarksnumbered=true,
            breaklinks=true,
            pdfpagemode=Fullscreen,
            pdfstartview=FitBH]{hyperref}
\usepackage[dotinlabels]{titletoc}
\usepackage{titlesec}

\definecolor{gesfpurple}{rgb}{0.47,0.19,0.42}
\newcommand{\gpurple}[1]{{\color{gesfpurple} #1}}
\definecolor{gesflanse}{rgb}{0.00,0.50,0.50}

\definecolor{gesfblue}{rgb}{0.08,0.42,0.76}

\definecolor{gesfred}{rgb}{1,0,0}

\definecolor{gesfwhite}{rgb}{1,1,1}

\definecolor{gesfblack}{rgb}{0,0,0}

\newcommand{\geqn}[1]{\hypersetup{linkcolor=blue}(\ref{#1})\hypersetup{linkcolor=blue}}
\newcommand{\gfig}[1]{{\hypersetup{linkcolor=violet}Fig.~\ref{#1}\hypersetup{linkcolor=blue}}}

\newcommand\pubnumber{}
\newcommand\pubdate{\today}

\def\Title#1{\begin{center} {\Large #1 } \end{center}}
\def\Author#1{\begin{center}{ \sc #1} \end{center}}
\def\Address#1{\begin{center}{ \it #1} \end{center}}

\newcommand\pubblock{\rightline{\begin{tabular}{l} \pubnumber\\
         \pubdate  \end{tabular}}}
\newenvironment{Abstract}{\begin{quotation}  }{\end{quotation}}
\newenvironment{Presented}{\begin{quotation} \begin{center} 
             PRESENTED AT\end{center}\bigskip 
      \begin{center}\begin{large}}{\end{large}\end{center} \end{quotation}}

\def\beq{\begin{equation}}
\def\eeq#1{\label{#1}\end{equation}}
\def\eeqn{\end{equation}}

\def\beqa{\begin{eqnarray}}
\def\eeqa#1{\label{#1}\end{eqnarray}}
\def\eeqan{\end{eqnarray}}

\let\bar=\overbar

\def\Dslash{\not{\hbox{\kern-4pt $D$}}}
\def\dslash{\not{\hbox{\kern-2pt $\del$}}}

\def\msb{{\bar{\ssstyle M \kern -1pt S}}}

\begin{document}
\begin{titlepage}
\pubblock

\vfill
\Title{Measuring the Leptonic Dirac CP Phase with TNT2K}
\vfill
\Author{Shao-Feng Ge \footnote{gesf02@gmail.com}}
\Address{Max-Planck-Institut f\"{u}r Kernphysik, Heidelberg 69117, Germany}
\vfill
\begin{Abstract}
I describe how the TNT2K (Tokai and Toyama to Kamioka) configuration with a
muon decay at rest ($\mu$DAR) add-on to T2(H)K can achieve better
measurement of the leptonic Dirac CP phase $\delta_D$. It has
five-fold advantages of high efficiency, smaller CP uncertainty, absence
of degeneracy, as well as guaranteeing CP sensitivity against
non-unitary mixing (NUM) and non-standard interaction (NSI). In comparison
to the flux upgrade with T2K-II, the detector upgrade with T2HK, and the baseline
upgrade with T2HKK, TNT2K is a totally different concept with spectrum upgrade
to solve the intrinsic problems in current and next generations of CP measurement
experiments. With a single $\mu$DAR source, TNT2K is much cheaper and technically
much easier than the DAE$\delta$ALUS proposal. The latter needs three sources that cannot
run simultaneously and consequently requires much higher fluxes. The single
$\mu$DAR source at TNT2K also allows a single near detector ($\mu$Near) to
fully utilize the neutrino flux for the purpose of constraining NUM,
but this is impossible at DAE$\delta$ALUS with three spatially separated sources.
\end{Abstract}
\vfill
\begin{Presented}
NuPhys2016, Prospects in Neutrino Physics\\[1mm]
Barbican Centre, London, UK,  December 12--14, 2016
\end{Presented}
\vfill
\end{titlepage}
\def\thefootnote{\fnsymbol{footnote}}
\setcounter{footnote}{0}

\section{The Intrinsic Problems in CP Measurement}

Both T2K and NO$\nu$A measure $\delta_D$ by observing the $\nu_\mu \rightarrow \nu_e$
oscillation. To maximize event rate, the neutrino energy and baseline are matched
to put oscillation at the first peak, reducing
the oscillation probability to have only $\sin \delta_D$ dependence,
\vspace{-1mm}
\begin{equation}
  P_{\overset{\nu_\mu \rightarrow \nu_e}{\bar \nu_\mu \rightarrow \bar \nu_e}}
\approx
  4 s^2_a c^2_r s^2_r
\mp
  8 c_a s_a c^2_r s_r c_s s_s \sin \phi_{21}
  \sin \delta_D \,,
\label{eq:Pme}
\vspace{-1mm}
\end{equation}
where
$(\theta_a, \theta_r, \theta_s) \equiv (\theta_{23}, \theta_{13}, \theta_{12})$,
$(c_x, s_x) \equiv (\cos \theta_x, \sin \theta_x)$, and
$\phi_{ij} \equiv \delta m^2_{ij} L / 4 E_\nu$. 

The feature of only $\sin \delta_D$ dependence causes several intrinsic problems.
First, the CP term has opposite sign in the neutrino and anti-neutrino modes.
With relative suppression by $c_s s_s \sin \phi_{21} / s_r \approx 1/5$, the
CP term can be easily smeared by the uncertainty of $s^2_a$ in the first term
of \geqn{eq:Pme}. Fortunately, we can extract $\sin \delta_D$
from the difference
$P_{\nu_\mu \rightarrow \nu_e} - P_{\bar \nu_\mu \rightarrow \bar \nu_e}$ by measuring
both neutrino and anti-neutrino modes. To gather comparable event rates, the
anti-neutrino mode needs much more time than the neutrino mode due to
smaller cross section, $\sigma_{\bar \nu} < \sigma_\nu$.
Roughly speaking, the anti-neutrino mode requires at least $2/3$ of run time.
This significantly reduces the event rates, leading to efficiency
problem.
Secondly, extracting $\sin \delta_D$ cannot uniquely determine $\delta_D$, with
degeneracy between $\delta_D$ and $\pi - \delta_D$.
Thirdly, the CP uncertainty is proportional to $|1/\cos \delta_D|$
with only $\sin \delta_D$ dependence. The recent global fits with preference
for maximal CP $\delta_D \approx - \pi/2$ is not good news for precision measurement.

These three problems are intrinsic problems for accelerator-based experiments,
including T2K, NO$\nu$A, and the future DUNE. In addition, T2K-II and T2HK have
exactly the same configuration and hence the same problems.
The baseline upgrade T2HKK seems to have better chance. However, it needs to
sit at the second oscillation peak to maximize event rate, still leading to the
same problems.

\section{CP Measurement at TNT2K}

TNT2K is designed for better CP measurement by supplementing T2K (T2HK) with a
$\mu$DAR source \cite{TNT2K} close to SK (HK). This requires a cyclotron to produce
the $\mu$DAR neutrino flux by accelerating proton to hit target, producing charged
pions which decay through the chain $\pi^+ \rightarrow \mu^+ + \nu_\mu
\rightarrow (e^+ + \nu_e + \bar \nu_\mu) + \nu_\mu$.
The $\bar \nu_\mu \rightarrow \bar \nu_e$ channel can be measured via inverse beta decay,
$\bar \nu_e + p \rightarrow e^+ + n$, with double coincidence.
Then T2(H)K can devote all time to the neutrino mode while $\mu$Kam,
$\mu$SK ($\equiv \mu$DAR+SK) or $\mu$HK ($\equiv \mu$DAR+HK), measures the
anti-neutrino mode. With much larger flux and
shorter baseline, $\mu$Kam can collect much more anti-neutrino events than T2(H)K.
This combination can significantly improve the efficiency by a
factor of 3 (4) in the neutrino (anti-neutrino) mode \cite{NSI}.
\begin{equation*}
\setlength{\tabcolsep}{3mm}
\centering
\begin{tabular}{c|cccc}
  $\delta^{true}_D = -90^\circ$ & T2K & $\mu$SK & T2K+$\mu$SK & $\nu$T2K+$\mu$SK \\
\hline
  Event Numbers & $\gpurple{\bf 114 \nu + 56 \bar \nu}$ & $\gpurple{\bf 212 \bar \nu}$ & $\gpurple{\bf 57 \nu + 268 \bar \nu}$ & $\gpurple{\bf 342 \nu + 212 \bar \nu}$ \\
\end{tabular}
\end{equation*}

\begin{figure}[h!]
\centering
\vspace{-2mm}
\includegraphics[width=4.5cm,height=7cm,angle=-90]{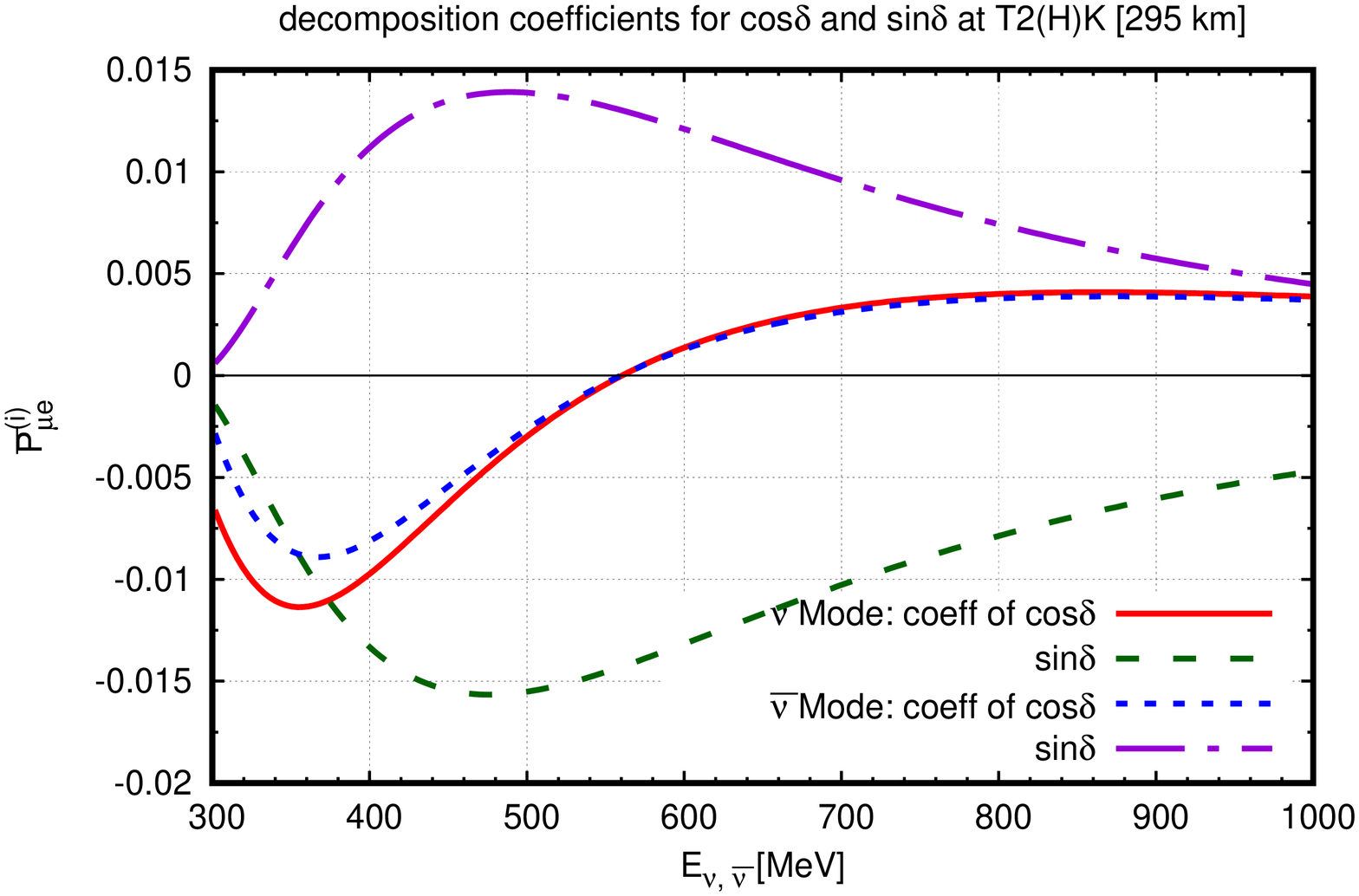}
\includegraphics[width=4.5cm,height=7cm,angle=-90]{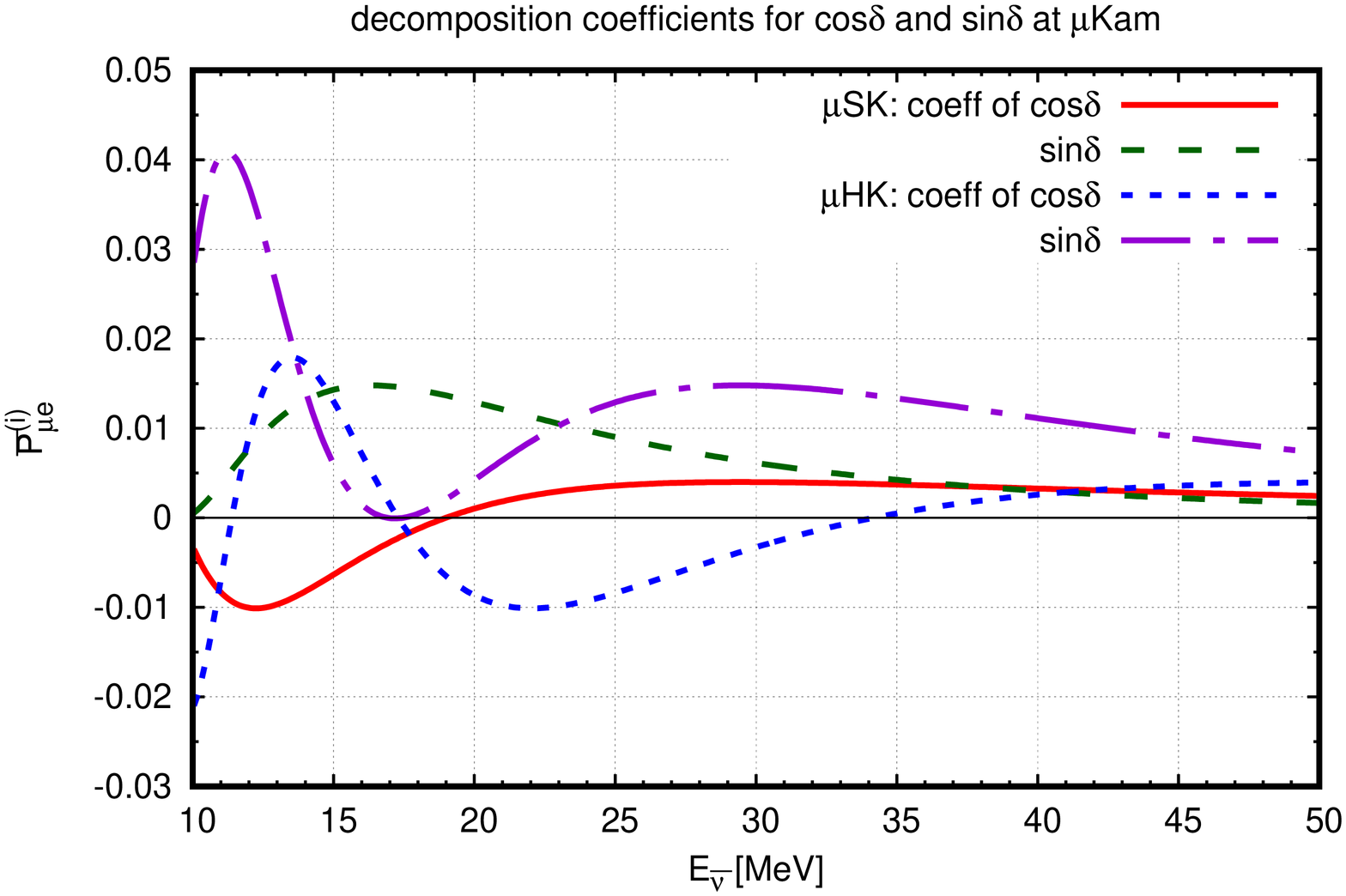}
\vspace{-3mm}
\caption{The decomposed coefficients of $P_{\mu e}$ and $\bar P_{\mu e}$
at T2(H)K and $\mu$S(H)K.}
\label{fig:deCoeff}
\end{figure}

\begin{figure}[h]
\centering
\includegraphics[width=5.9cm,height=0.7\textwidth,angle=-90]{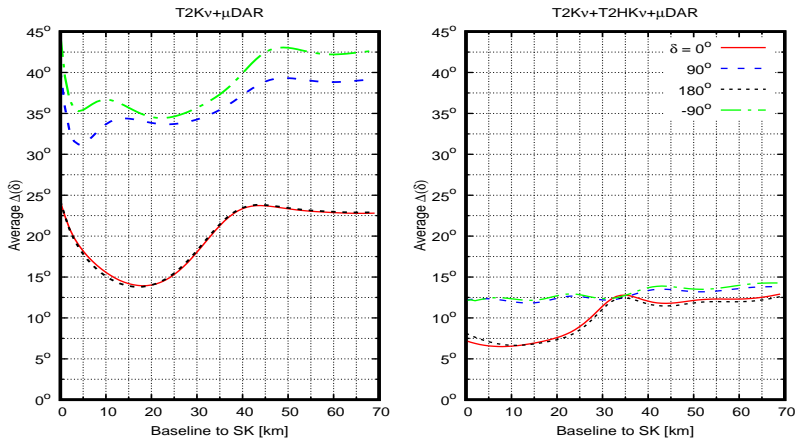}
\caption{The baseline dependence of CP sensitivity at TNT2K without or with HK.}
\label{fig:CP}
\end{figure}

The $\bar \nu_\mu$ flux is produced from $\mu^+$ decay at rest with
a wide and flat spectrum across the interested energy
range, $30\,\mbox{MeV} \lesssim E_\nu \lesssim 55\,\mbox{MeV}$.
Using the decomposition formalism \cite{Decomposition} in the
propagation basis in \gfig{fig:deCoeff},
we can clearly see vanishing $\cos \delta_D$ term at the J-PARC spectrum peak,
$E_\nu \approx 600\,\mbox{MeV}$, in contrast to comparable $\cos \delta_D$
and $\sin \delta_D$ terms across the $\mu$DAR energy range, allowing TNT2K to
avoid degeneracy and large uncertainty problems with the help of $\cos \delta_D$
dependence.
\gfig{fig:CP} shows how CP uncertainty depends on baseline with optimal distance
around $23\,\mbox{km}$ \cite{TNT2K}.

\section{Non-Unitary Mixing}

If heavy neutrino exists and mix with light neutrinos,
the usual $3 \times 3$ light neutrino mixing matrix becomes non-unitary,
 \begin{equation}
   N
 =
   N^{NP} U
 =
 \left\lgroup
 \begin{array}{ccc} 
 \alpha_{11} & 0 & 0\\
 \alpha_{21} & \alpha_{22} & 0\\
 \alpha_{31} & \alpha_{32} & \alpha_{33}
 \end{array}
 \right\rgroup U \,.
 \label{eq:N}
 \end{equation}
For $\mu \rightarrow e$ transition, the phase in
$\alpha_{21} \equiv |\alpha_{21}| e^{- i \phi}$ can fake CP effect,
\begin{eqnarray}
   P^{NP}_{\mu e}
 & \hspace{-3mm} = \hspace{-3mm} &
   \alpha^2_{11}
 \left\{
   \alpha^2_{22}
 \left[
   c^2_a |S'_{12}|^2
 + s^2_a |S'_{13}|^2
 + 2 c_a s_a (\cos \delta_D \mathbb R - \sin \delta_D \mathbb I) (S'_{12} S'^*_{13})
 \right]
 + |\alpha_{21}|^2 P_{ee}
 \right.
 \nonumber
 \\
 & \hspace{-3mm} + \hspace{-3mm} &
 \left.
   2 \alpha_{22} |\alpha_{21}|
 \left[
   c_a \left( c_\phi \mathbb R - s_\phi \mathbb I \right) (S'_{11} S'^*_{12})
 + s_a \left( c_{\phi + \delta_D} \mathbb R - s_{\phi + \delta_D} \mathbb I \right) (S'_{11} S'^*_{13})
 \right]
 \right\} \,.
 \label{eq:PmeNP}
 \end{eqnarray}
In addition to $(\cos \delta_D, \sin \delta_D)$, four extra CP terms
$(c_\phi, s_\phi)$ and $(c_{\phi + \delta_D}, s_{\phi + \delta_D})$ appear.
The CP sensitivity at T2(H)K can be significantly reduced.
\begin{figure}[h]
\centering
\includegraphics[height=0.3\textwidth,angle=-90]{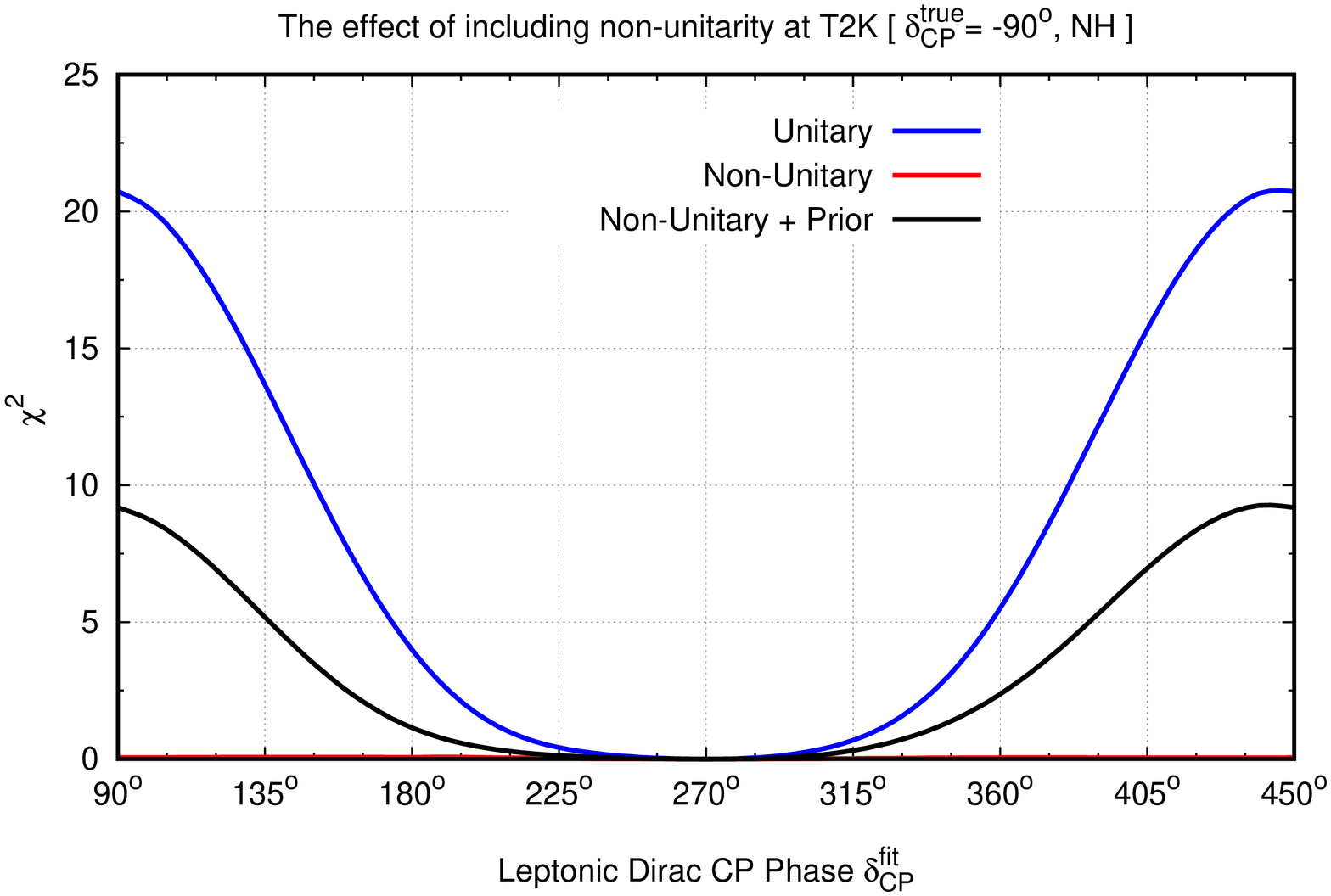}
\includegraphics[height=0.3\textwidth,angle=270]{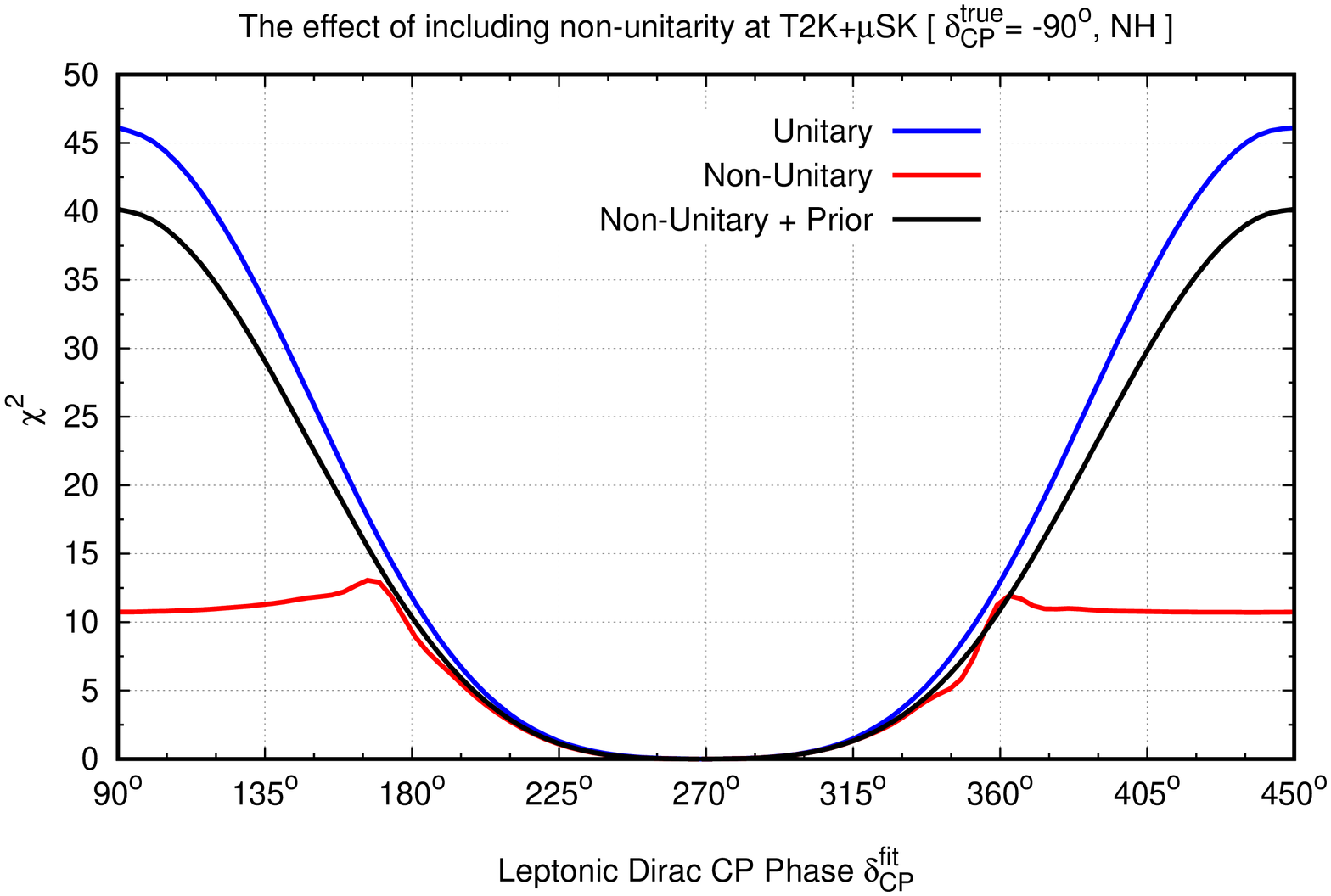}
\includegraphics[height=0.3\textwidth,angle=270]{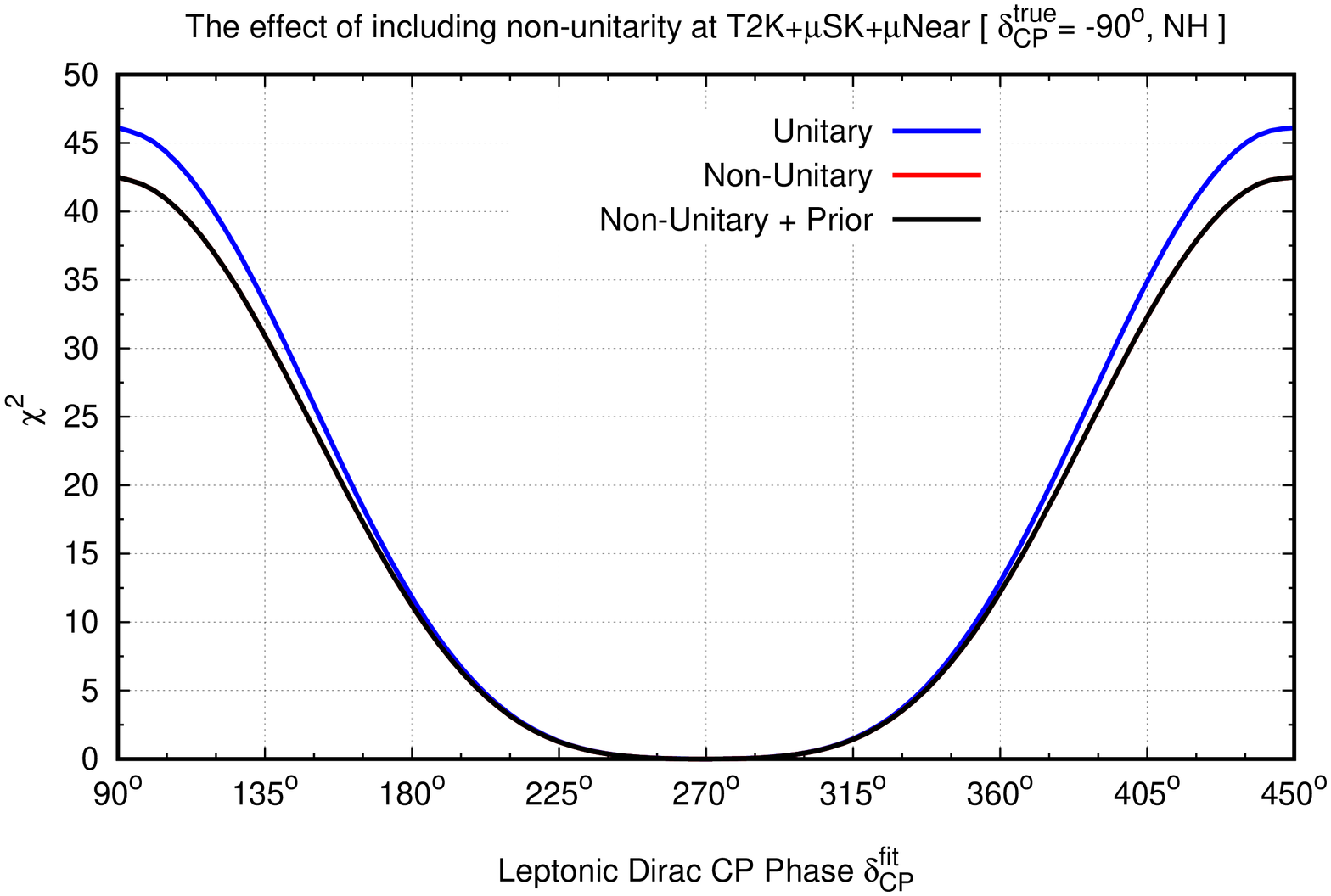}
\caption{The CP sensitivities at T2K, T2K+$\mu$SK, and T2K+$\mu$SK+$\mu$Near.}
\label{fig:NUM}
\end{figure}
The TNT2K configuration can partially improve the situation due to the presence of
$\cos \delta_D$ dependence. Further adding a near detector
close to the $\mu$DAR source can fully restore the CP sensitivity by measuring
the {\it zero-distance effect}, 
$\left. P^{NP}_{\mu e} \right|_{L \rightarrow 0} \rightarrow |\alpha_{21}|^2$,
to constrain the size $|\alpha_{21}|$ of the extra CP term \cite{NUM}.

\section{Non-Standard Interaction}

The CP effect can also be faked by NSI. Since NSI enters oscillation
as matter potential, its effect is proportional to the neutrino energy
which is unfortunately not small at T2K and NO$\nu$A.
\begin{figure}[h]
\centering
\includegraphics[width=3.9cm,height=0.45\textwidth,angle=-90]{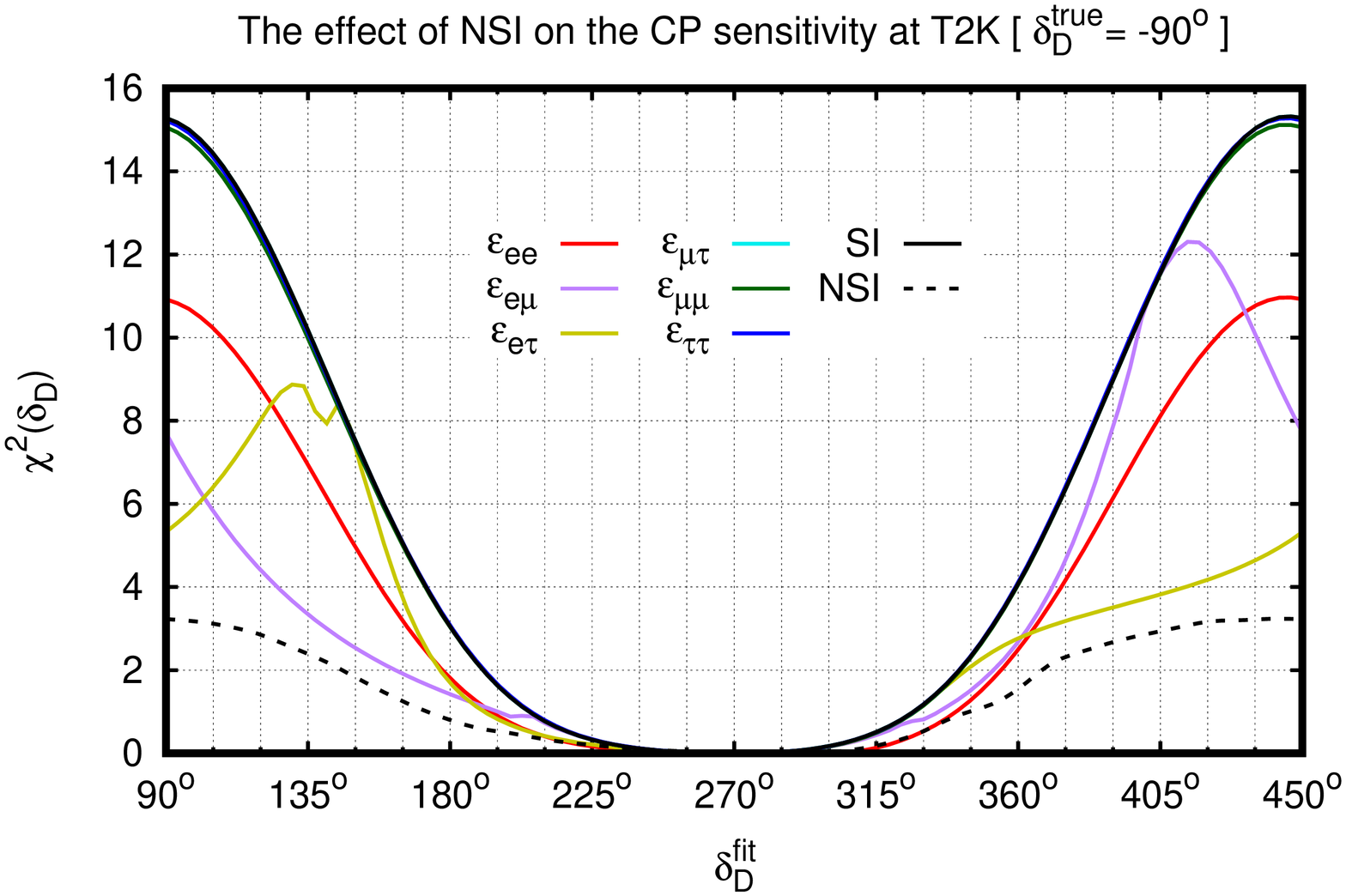}
\includegraphics[width=3.9cm,height=0.45\textwidth,angle=-90]{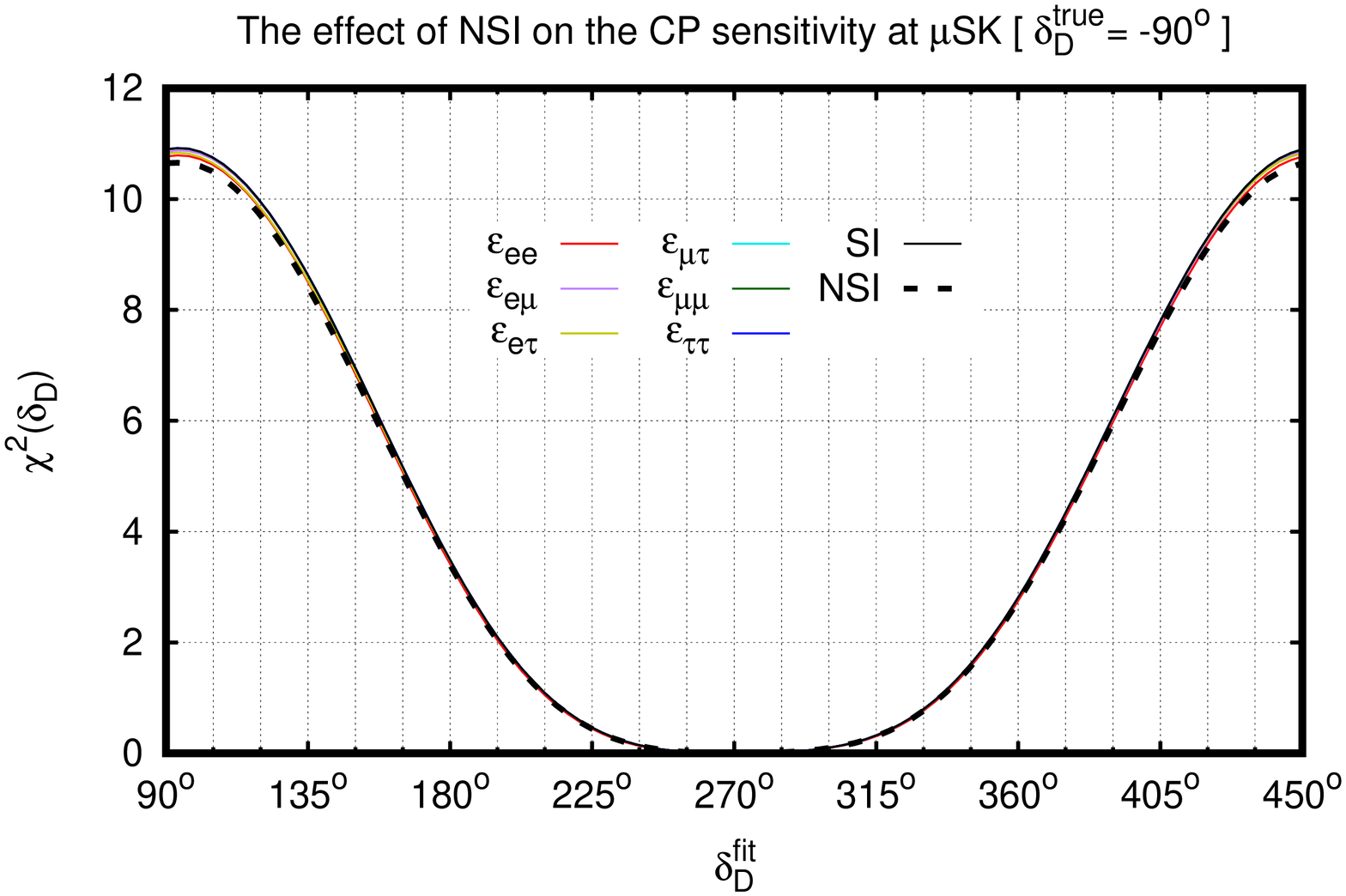}
\\
\includegraphics[width=3.9cm,height=0.45\textwidth,angle=-90]{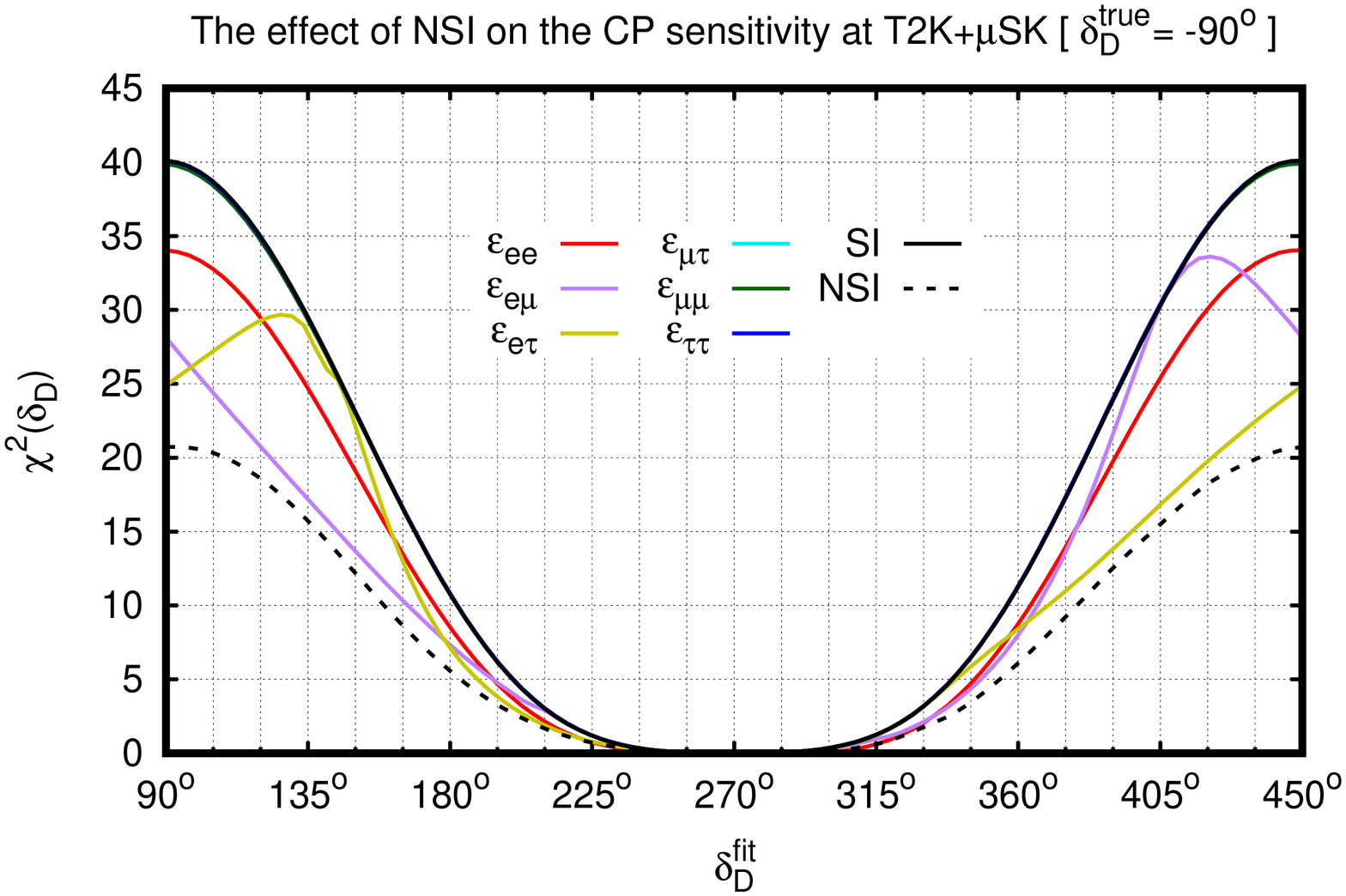}
\includegraphics[width=3.9cm,height=0.45\textwidth,angle=-90]{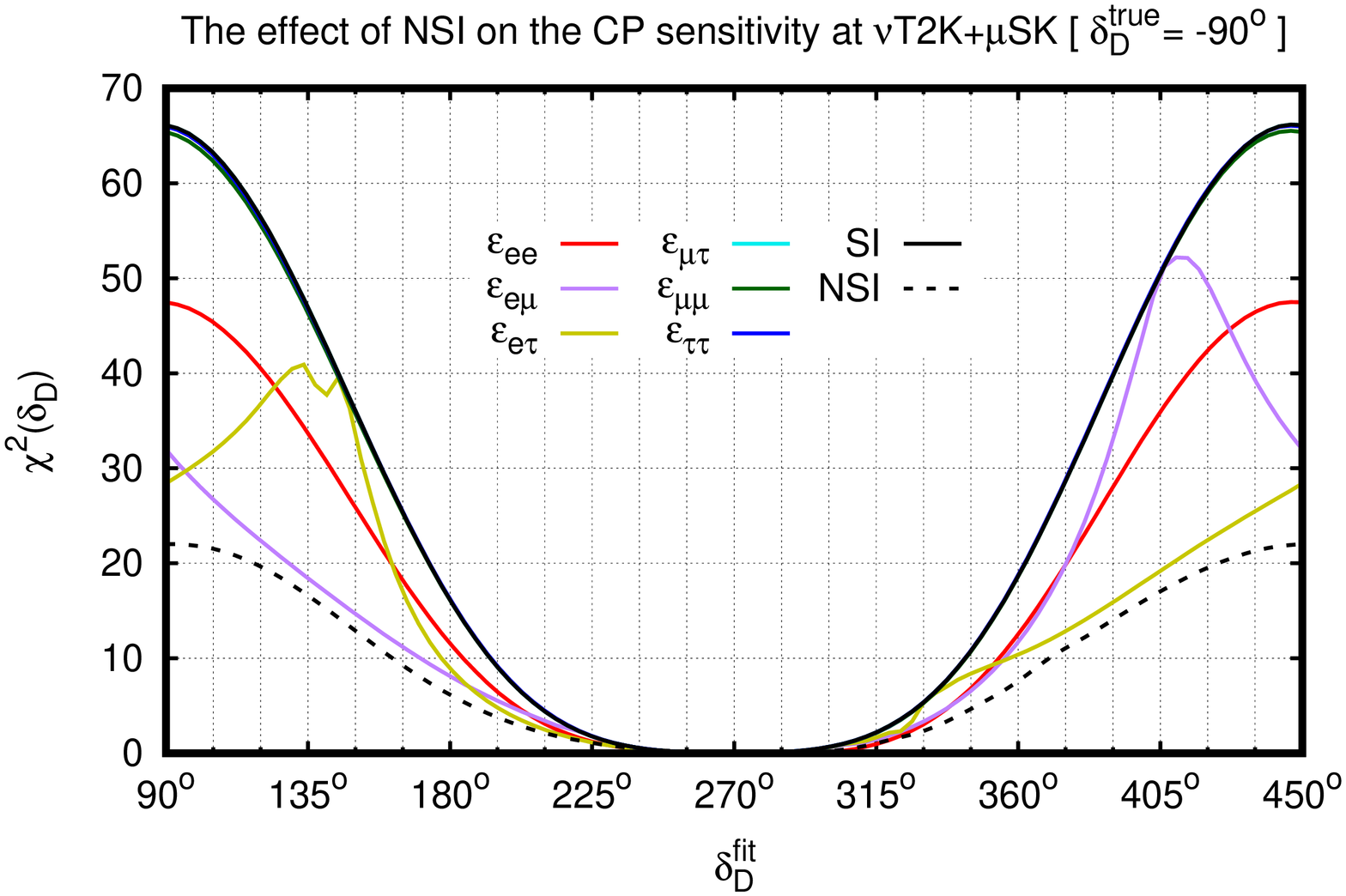}
\caption{The CP sensitivity in the presence of NSI.}
\label{fig:NSI}
\end{figure}
As show in the first subplot of \gfig{fig:NSI}, the CP sensitivity at T2K
$\Delta \chi^2 \approx 15$
can be significantly reduced by a factor of $5$. In contrast, the neutrino energy
of the $\mu$DAR flux is smaller than T2K by a factor of $10$ and feels negligible
effect from NSI, see the second subplot of \gfig{fig:NSI}.
While T2K measures both the genuine CP $\delta_D$ and NSI,
$\mu$DAR focuses on $\delta_D$. As shown in the last two subplots in \gfig{fig:NSI},
TNT2K can measure $\delta_D$ and NSI simultaneously, hence guaranteeing the
CP sensitivity against NSI \cite{NSI}.

\section{Comparison with DAE$\delta$ALUS}

Requiring only one cyclotron, TNT2K can run with duty factor close to $100\%$,
in contrast to the $25\%\sim30\%$ of DAE$\delta$ALUS. The later needs 3 $\mu$DAR
sources but they cannot run simultaneously. Otherwise, it is impossible
to tell from which source the neutrinos come from and how long they have traveled.
To achieve the
same $\mu$DAR event rate as TNT2K, DAE$\delta$ALUS demands much higher flux
which is inversely proportional to duty factor
and hence more advanced technology. In addition, the $\mu$Near detector for
constraining NUM can use the full $\mu$DAR flux
at TNT2K but this is impossible at DAE$\delta$ALUS with distributed sources.
Comparing with DAE$\delta$ALUS, TNT2K is cheaper with only one cyclotron,
technically easier with lower flux, and physically has more potential.

\end{document}